\documentclass{article}

\usepackage{url,hyperref,microtype,subcaption} 
\usepackage[onehalfspacing]{setspace}
\usepackage{tabularx}
\usepackage{arydshln}
\usepackage{natbib}
\usepackage{graphicx}
\usepackage[dvipsnames]{xcolor}

\def\Authors{Martin Cooney\,$^{1,*}$}

\begin{document}
\onecolumn

\textcolor{red}{Please note: This is a preprint version of a journal article before reviews submitted on 2020-5-7 to The Art of Human-Robot Interaction: Creative Perspectives from Design and the Arts, hosted by Dr(s) Damith C Herath, Elizabeth Ann Jochum, Christian Kroos, and David St-Onge in Frontiers in Robotics and AI - section Human-Robot Interaction.}

\title[Personalized Robot Art]{Robot art, in the eye of the beholder?: Personalization through self-disclosure facilitates visual communication of emotions in representational art} 

\author[]{\Authors}


\maketitle

\begin{abstract}


Socially assistive robots could help to support people's well-being in contexts such as art therapy where human therapists are scarce, by making art such as paintings together with people in a way that is emotionally contingent and creative.
However, current art-making robots are typically either contingent, controlled as a tool by a human artist, or creative, programmed to paint independently, possibly because some complex and idiosyncratic concepts related to art, such as emotion and creativity, are not yet well understood.
For example, the role of personalized taste in forming beauty evaluations has been studied in empirical aesthetics, but how to generate art which appears to an individual to express certain emotions such as happiness or sadness is less clear. 
In the current article, a collaborative prototyping/Wizard of Oz approach was used to explore generic robot art-making strategies and personalization of art via an open-ended emotion profile intended to identify tricky concepts.
As a result of conducting an exploratory user study, participants indicated some preference for a robot art-making strategy involving ``visual metaphors'' to balance exogenous and endogenous components, and personalized representational sketches were reported to convey emotion more clearly than generic sketches.
The article closes by discussing personalized abstract art as well as suggestions for richer art-making strategies and user models.
Thus, the main conceptual advance of the current work lies in suggesting how an interactive robot can visually express emotions through personalized art; the general significance is that some positive feedback has been obtained toward this direction toward an interactive scenario involving emotions and creativity, which could inform next steps and potentially also be incorporated into robots in everyday human environments, toward providing extra value for and facilitating technological acceptance.

\end{abstract}

\section{Introduction}

Will interactive social robots start to be used in everyday places, and if so, what will they be like?
There seems to be a gap between recent failures to market such robots and expectations that such robots could help us to feel good and flourish \citep{dereshev2018smart}.
On the one hand, the world has possibly never been as ready for the integration of social robot technologies into domestic and public human environments:
Various dreams of social robots as our dependants, servants, teachers, companions, partners, and guardians have been shared across the world through fictional works like Astro Boy (1952) \citep{nakao2014robots}, The Jetsons (1962), Doraemon (1970) \citep{marshall2016doraemon}, Star Wars (1977), Short Circuit (1986), Cherry 2000 (1987), and Terminator 2 (1991). 
Simultaneously, our world has been ``updated'' by numerous technological innovations such as the internet, inexpensive robot toys such as drones, home voice assistants, and smartphones; the recent coronavirus pandemic has also led to a general decrease in close contact with others, which could be supplemented by robots.
What then might be missing?
Here the current article puts forward two conjectures:
\begin{itemize}
\item{One, that social robots will follow the path of smartphones in becoming prevalent once they are perceived to provide multiformed value in excess of cost.
Where a smartphone can be used for many useful tasks like telling time, predicting weather, and playing media, robots will assist activities of daily living, like cleaning, fetching, cooking, folding clothes, washing dishes, as well as socially interact with us by playing with us (perhaps hugging \citep{block2019softness} or even biting us! \citep{nakagawa2020effect}), reporting on how health by collating heterogeneous data and monitoring unseen signals, helping in emergencies, and providing social-physical therapy such as exercise instruction \citep{fitter2018teaching}, inter alia.
To realize such a scenario, activities could include trying to work from the ground up by finding behaviors which could be added to existing home voice assistants, meliorating performance in established tasks (and also possibly making code available via common frameworks such as Robot Operating System)--and exploring some new applications, as the current article seeks to do.
Here this is referred to as the  ``smartphone hypothesis'' for social robots.}
\item{Two, that one useful category of tasks to investigate involves emotions and creativity\citep{picard1995affective,colton2012computational}, such as art-making. 
Emotions and creativity, as defined in Table \ref{tabDefinitions}, are typically associated with humans. But, given that many of our past preconceptions have been dispelled, such as that artificial intelligence could not play complex games like Go, this line of research could one day affect how we think about emotions and creativity, and even ourselves (e.g., if emotions and creativity can no longer be used as a \textit{differentia specifica} for humans, what might be next for Plato's ``featherless biped''?).
Moreover, although the idea of robots which can make art has fascinated people for a long time \citep{herath2016robots}, there has recently been renewed interest in robot art as a means to explore new ways of thinking about interactive robots \citep{st2019robotic}.
Art-making is also an activity which people of all ages and cultures can enjoy, where painting together with others can positively affect a person's restfulness, self-image, stress tolerance, and vital signs--facilitated by processes of self-exploration, self-fulfillment, catharsis, and perceiving belonging \citep{stuckey2010connection}.
From a therapeutic perspective, work in this line of research can also help to address a social need related to a growing shortage of human caregivers (e.g., persisting loneliness has been described as a growing problem linked to high cost and ruinous health outcomes, which might be exacerbated by isolation caused by the coronavirus).
Some ongoing work has also involved the first testing of an art-making robot at a hospital, resulting in some positive initial feedback \citep{herath2020arts+}.}
\end{itemize}

\begin{figure}[h!]
\begin{center}
\includegraphics[width=\linewidth]{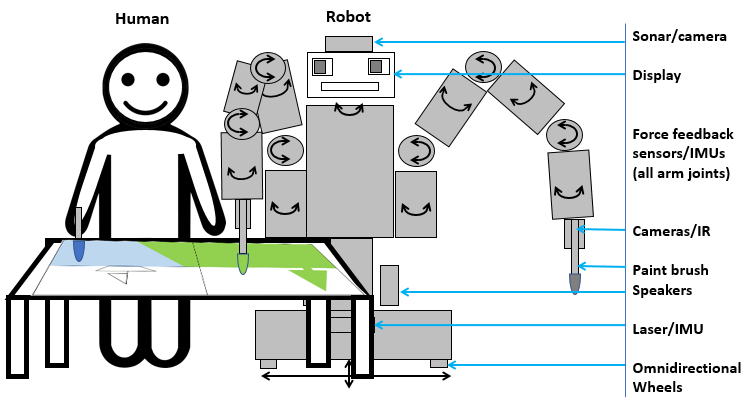}
\end{center}
\caption{Basic concept. A social robot can be designed to interact with a human in various ways, including emotional and creative tasks such as art-making that provide enjoyment or therapeutic value.}\label{figBasicConcept}
\end{figure}

Given this complex and challenging scenario, how can the design space for such an interactive robot be explored?
One broad criticism of social robotics is that it has been constrained by normative demands, whereas somaesthetic experiences are highly idiosyncratic \citep{kerruish2017affective}. 
In other words, both general insight into basic underlying patterns and codes required to get started, as well as an approach for modeling personal differences should be considered.
Such knowledge has to some extent been elucidated in the related field of empirical aesthetics in regard to beauty evaluations--for example, studies have reported on how private and shared taste influence human perception depending on the degree to which symbols are abstracted \citep{leder2016private}--but how a robot can generate art which appears to express certain emotions such as happiness or sadness to an individual is less clear.
Therefore, the current article investigates how to design a socially assistive painting robot which uses a personalized model to visually convey emotions, as illustrated in Figure \ref{figBasicConcept}--this involves exploring a general art-making strategy, as well as how to build and apply such a personalized model, also considering the degree to which art is abstracted.

To achieve this goal, a basic scenario was identified and explored through a mid-fidelity collaborative prototyping approach involving some artists.
The contribution of the current work is insight into some questions which arose in prototyping, obtained by conducting a simplified user study using the Wizard of Oz approach to generate art based on modeling individual perceptions, which is intended to inform next steps in this area.

\begin{table}
\caption{Some definitions of terms used in this article.}
\label{tabDefinitions}  
\begin{tabularx}{\linewidth}{ | >{\hsize=0.2\hsize}X | >{\hsize=0.8\hsize}X | } \hline

Socially assistive robot	 & An embedded computing system, comprising sensors and actuators which afford some semi-autonomous, intelligent, or human-like qualities, intended to interact to support people's well-being  \\  \hline
Well-being	& A subjectively perceived state, related to happiness, life satisfaction, and quality of life, encompassing physical, psychological, and social factors (hedonic and eudaimonic), and linked with positive emotions and creativity. \\  \hline
Emotion	& A complex psycho-physical phenomenon involving cognitive appraisals, subjective feelings, somatic symptoms, and affect displays, related to sentiment and mood, which is typically encoded in a simplified manner via dimensions or categories in computers (``the affective gap''). One important emotional process is empathy, the capability to demonstrate recognition of and caring for another's emotions.\\  \hline
Creativity	& A way of doing things characterized by novelty, not something one has or doesn't have \citep{gershgorn2016can} \\ \hline
Personalization	& A process also referred to as customization, tailoring, or targeted servicing, which has been reported to have various positive effects, such as on engagement and trust  \citep{sillence2006framework}, especially in areas where people are highly different\\  \hline
Symbols &  Here symbol does not refer to symbolic art, which was a reaction against realism; it refers here to some representation of a concept, person, or thing\\ \hline
Abstract &  Abstraction can be present to various degrees in artwork; here ``abstract'' is used to mean nonrepresentational in the sense that people and objects cannot be clearly recognized--rather the art uses shapes and colors to evoke impressions \\  \hline
\end{tabularx} 
\end{table}

\section{Methods and Materials}
 
\subsection{Painting together with a human}

The current section summarizes some of our previous work, which involved first identifying a basic scenario informed by the related literature \citep{cooney2018design}.  
For simplicity, the basic scenario selected for exploration was dyadic (one human and robot), visual (non-verbal), over a single session (e.g., 10 min), and with free subject matter. Requirements for a robot included safety to interact near people, as well as a capability to paint (human-like reach and cameras), and a familiar interface which could support social interactions (such as a ``skeuomorphic'' humanoid form); based on these requirements, a Baxter robot was chosen. 
Regarding an art-making strategy, from a simplified perspective, two common kinds of robots which have been built to create art are referred to here as ``exogenous'' or ``endogenous'':
Exogenous robots are tools that are dependent on and controlled by humans, such that some physical signal by a human such as motion or sound results in some physical motion of the robot; an extreme example could be a photocopier. By contrast, endogenous robots mostly operate independently of a human artist in a non-interactive stand-alone fashion; an example could be a robot arm that is programmed to paint a specific scene. 

Based on this, a mid-fidelity research prototyping approach was followed to gain some insight\citep{cooney2019designing}.
Research prototyping and constructive design allow researchers to gain pragmatic insight into complex phenomena in the real world, where computer simulations and deduction would be insufficient for understanding. 
Mid-fidelity prototyping additionally provides a useful balance of accurate insight into how a finalized system might perform, while allowing results to be obtained quickly and practically.
Furthermore, collaboration between artists and engineers can have positive synergistic effects \citep{st2011comparison}.
Thus, some first exploration was conducted with two artists and some engineering students, which involved the artists trying different setups with the students controlling the robot. 
One idea was that the robot could paint in a contingent way to indicate empathy, where the simplest way to suggest that a robot understands what a human has painted is direct mimicking, or painting the same thing the human has painted. 
This felt interesting, in that the robot seemed to have some perceptive capability, and the human's art was attributed enough importance to mimic, but merely copying also felt like the robot was a mere machine and not like a partner.
Another enjoyable interaction involved an artist painting together, face-to-face, with the robot on a shared canvas, where the robot painted independently of the artist; although artistically interesting (due to the human artist's improvisation), the lack of a connection or bond also felt like a human working with a machine, not a partner. 
Thus, both strategies felt limiting due to their one-way nature, and led us to consider if a robot could produce art which balances exogenous and endogenous components to appear both empathetic and creative. Thus a robot could base its art partially on what the human is doing, and partially on its own intention (e.g., to help the human to feel happy).

Being able to both recognize what a human does, and produce something new based on this and the robot's own agenda, might indicate some degree of intelligence, as robots which are perceived as intentional agents with a mind (i.e., ``mind perception'') are more likely to be treated as a partner and be the recipients of empathy, morality or prosocial behaviors  \citep{wiese2017robots}. 
Also, by creatively reshaping a human's emotional expression, the art might be more likely to be stimulating, thought-provoking, meaningful, and facilitating self-exploration.
This approach additionally seemed to coincide with the ``responsive art'' approach in art therapy, in which a therapist provides visual feedback based on a person's art, which has been described as effective toward exploring the meaning of a person's art.
Moreover, the usefulness of a robot's behavior appearing to balance exogenous and endogenous components to indicate agency has also been previously suggested in HRI, within a related context of how a robot can direct attention toward interacting people \citep{kroos2012evoking}.

As a first step to explore such a strategy, the use of a ``visual metaphor'' is proposed here, loosely in line with the concept of the ``adjacent possible''. Namely, that a robot can paint something which is similar in emotional meaning but different in expression, based on our idea that different symbols can be painted to express the same emotion. For example, if a person paints something relaxing like a quiet forest, the robot could paint something else relaxing, like a babbling brook. Grieving people could be painted beside a remorseful grave scene. A snake poised to strike could be painted beside a threatening gun. A bright set of balloons could be drawn beside a festive stack of presents. This idea is general and can also be applied to abstract art. For example, if a person uses diagonal lines to express arousal, a robot could instead use an aroused warm color. Thus, the term ``metaphor'' here is used to describe such differing symbols which are intended to express similar emotional meaning. For example, a two-dimensional painted circle can be said to be an abstracted symbol representing a three-dimensional balloon; however, it might be useful to paint such a circle, not because there is any specific meaning in depicting a balloon, but rather as a metaphor to express an emotion of joy.

\begin{figure}[h!]
\begin{center}
\includegraphics[width=\linewidth]{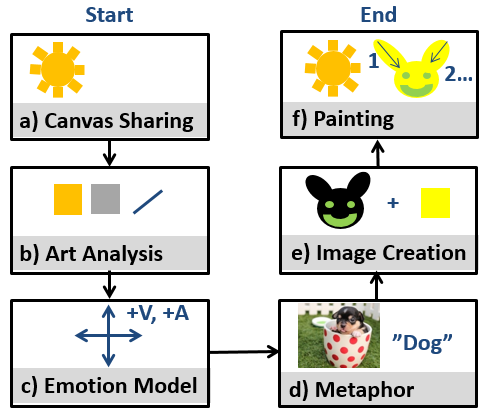}
\end{center}
\caption{A process flow for using visual metaphors to convey contingency and artistry in art that a robot creates with a human }\label{figBasicFlow}
\end{figure}

How then to incorporate this concept into an interaction?
As shown in Figure \ref{figBasicFlow}, a robot can detect when/where to paint; infer the emotional meaning of a person's art by analyzing colors, lines, and composition; select a contingent metaphor with an affective image database; generate an image based on the metaphor; and paint the image:

(a) Canvas sharing. It could be irritating for the robot to block, collide with, or otherwise interfere with the human when painting, or to require a human to compensate for the robot. Thus a robot should either be able to predict where a person will want to paint during the interaction and possibly have some logic to handle interruptions, or merely follow a turn-based interaction design. 
In the latter case, a user could indicate that it is the robot's turn to paint in various ways, each with some demerits. For example, it might be difficult for a person with a cognitive order to haptically control a robot, which is not required in typical interactions with humans. Controlling a robot verbally using CMU PocketSphinx or Google Speech might require the person to be able to speak with sufficiently clear enunciation and volume, and a strategy for dealing with the robot's own sounds (e.g. either speech or noise from the robot's actuators). Controlling a robot visually, e.g., through background subtraction with OpenCV, might require an environment with stable illumination and a strategy for dealing with changes due to the robot's own moving of sensors like cameras. Another alternative could be to combine various modalities for robustness.
For this simplified initial exploration, a turn-based design with haptic control was implemented, in which the user presses a button on the robot's arm to indicate it is the robot's turn to paint. (A visual system using the robot's arm camera was also initially built, but there were sometimes false detections due to shadows or flickering lights.)

(b) Physiological signal detection. 
To seem contingent, the robot should perceive what the human has expressed. Given that paintings are primarily visual, a camera, either on the robot or in the room, can be used to analyze a person's painting. Algorithms will likely become increasingly capable of high-level summarization and observation; currently, low-level and mid-level analysis is typical, whcih involves detecting colors, lines/shapes, composition, and depicted objects. The art that a person has drawn is not the only way to detect physiological signals which can be linked to emotion; in our previous work, a Brain-Machine Interface was used, and numerous other possibilities exist.

For our prototype conversion of an RGB image to HSV (Hue-Saturation-Value) space was conducted to color-pick six basic hues and calculate their average intensity, while using the Hough Transform to detect lines.

(c) Emotion Inference. Next the robot should seek to infer the emotional meaning underlying the physiological signals detected, such as visual expression or brainwaves. A personalized profile with information on how a person associates art with emotions could be used if available, or a generic model based on some emotion model. 
For our initial prototype, some simplified heuristics related to the visual arts were used in conjunction with a generic dimensional model of emotion: A linear combination of features based on the detected color and lines was used to calculate valence and arousal, for which Ståhl's model was used to calculate a contribution of each hue by area. Intensity also influenced valence (light being positive and dark being negative), and the incidence of diagonal lines affected arousal. 

(d) Metaphor Selection. Next the robot should seek to find a new different way to express the detected emotional meaning. Nouns with a similar corresponding average emotion could be looked up via a large sentiment lexicon such as Affective Norms for English Words (ANEW), SentiWord-Net, or WordNet-Affect, possibly with a concreteness rating to ensure that the noun can be expressed visually in a recognizable way, or an affective image database could be used.
A challenge identified was monosemy, which also related to bias and variance in data: not all images of dogs will induce the same emotional responses.

For our prototype, a simplified capability was implemented to search affective image databases; for example, looking for happy, relaxed, sad, and angry emotions resulted in images showing dogs, flowers, gray yarn, and injuries (OASIS), and skydiving, nature, a cemetery, and mutilation (IAPS), respectively.

(f) Image Generation (Reification). 
The next step is to move from a metaphor like ``dog'' to a specific image of a dog to paint.
For example, Deep Convolutional Generative Adversarial Networks (DC-GANs) such as StyleGAN, in conjunction with a large dataset of dog images from Google Image Search or ImageNet, could be used to generate a ``unique'' image of a dog. This image could then be abstracted, and the color and lines modified to more clearly convey an emotion, or to less resemble previous images that the robot has drawn.

For our prototype, a DC-GAN was used to develop compositions, but problems related to time were encountered, which prohibited the kind of online interaction we wished to achieve: web-scraping required much time; even with small MNIST-sized grayscale images, the algorithm took hours to generate new images; and with only hours of training the algorithm did not function automatically (a human was needed to select appropriate images from the output).
An alternative for initial exploration could involve asking the artists to come up with images for some typical metaphors.

(g-h) Painting Plan and Execution. 
Next the robot should map the image to paint into motions to execute to paint strokes on the canvas. This part of the process, although challenging like the other parts, is the one which has possibly seen the most focus from previous work (e.g., typically a visual feedback loop is used to compare the current canvas and desired target, while adding strokes \citep{lindemeier2015hardware}; it is also possible to simplify complicated images before depicting them, like in the work of Wang and colleagues, who used a CycleGAN and genetic algorithm for image-to-image translation \citep{wang2019robocodraw}).
An advanced algorithm might seek to also factor time into account and only conduct a subset of the most important strokes before giving the human another turn to avoid long waiting times. 
For our prototype, our artists were asked to generate some motions for the robot to perform based on images they created. Additionally, it was explored how the prototype could also seek to show emotions through a face in its display, as well as motions (curvature and velocity), and voice, during painting.

This process can be repeated for multiple turns, thereby closing the ``affective loop'' (perceiving emotion, acting, checking the result, and refining). 
A video describing this initial prototype is available online (https://www.youtube.com/watch?v=yeNXg2UKA-8).

\subsection{Personalization: User Study}

In the previous section, a generic strategy was described, but personalization is also essential since the perception of emotions in art is idiosyncratic.
For example, to some, artwork like Van Gogh's Starry Night might seem dreamy or despairing; Bosch's triptych might seem fascinating or frightening; Dali's depiction of memory melting like discarded cheese might feel resigned, or regretful in the vein of Shelley's Ozymandias; and a dark painting might be heartbroken, or happy for a member of the goth subculture--correctly identifying the meaning of a person's art could improve rapport, whereas mistakes could reduce the trust a person has in a robot, especially in a therapy setting.
Such effects have also been seen in practice; for example, personalization of a robot in a language class resulted in more positive emotions than a non-personalized system \citep{gordon2016affective}.
But what does this process typically involve?

Personalization is a common technique that involves applying knowledge of users to offer improved services which can be easier to use, more satisfying, and more persuasive.
It can be seen inter alia on the web in recommendations, advertisements, search results, and social media content; in interactive learning systems, through knowledge tracing;  in product development and testing in the form of data-driven personas; and in commercial products such as mugs, shirts, books, and statues that use photos, names, or 3D data.
Typically personalization in the digital realm involves user models and profiles: a user model describes how a user can be represented in terms of some properties of interest, like name or gender, which is used to structure a user profile, the data for a specific user or group.
Data are typically obtained explicitly in a user-driven manner by directly asking users, or implicitly in a system-driven manner, with various benefits and demerits; explicit user-driven approaches can overwhelm users, who also might not know exactly what they want from the start, whereas system-driven approaches can incorporate limiting hidden biases.

Some concerns for defining a user model include what properties to include and how to structure them, how the data will be obtained, and how the data can be used.
There seem to be many possible properties which could affect the perception of emotions in visual art, and commensurately many possible model configurations.
A naive or brute force method might seek to obtain data for every single concept that could be expressed through art from a person, which is not likely to be practical.
At a slightly higher level of abstraction--the stereotypical layer--properties which were felt to be important for the emotional rating of art in the OASIS affective image dataset included age, gender, geographic location, race, ideological self-placement, and socioeconomic background (highest level of education, and household income) \citep{kurdi2017introducing}. 
For communication on social media Zhao et al. proposed taking into account also the social context, the person's previous emotional state, and influence of location (e. g., a photo taken in entertainment venue might be happy, whereas a photo taken at a funeral might be sad), as well as personality via the Big Five model \citep{zhao2016predicting,zhao2019personalized}.
Similarly, Rudovic et al. proposed a model to encode how autistic children show emotions in their audiovisual and physiological behavior, at three levels: culture, gender, and individual.
 \citep{rudovic2018personalized}, but it's not clear if this is the only possible configuration: as a simple example, could the first layer be gender rather than culture, and what happens when other variables like age are considered? 
As well, how to move from concepts to a specific image to paint in a transparent way is also not clear from the literature.

As a simplified starting approach, the current article proposes a sparse and flat Folksonomic-style model in which open questions are used to capture ``tricky'' concepts which exert a large effect on individual perceptions.
A Folksonomic-style model can be useful in complex situations where it might be unclear how to construct a full taxonomy, or what kind of architecture would be appropriate.
Since there is no real way to know what someone is thinking without asking them, a user-driven self-disclosure is suggested, which could be more accurate than circumstantial guessing.
To avoid overwhelming users, the number of questions is restricted. As well, for simplicity, single emotions are considered.
To explore mapping from emotions to paintings, the current article suggests combining results from look-ups in an affective image database, followed by post-processing.
Thus, the part of the system related to the user model should accept a user identity and target emotional state as input, and output a paintable image expected to elicit this emotional state in the identified user.

Adding the conjecture from the previous section, the current article posits (1) that people would rather interact with a robot that is both contingent and creative, rather than just one or the other, and (2) that personalized art based on self-disclosure could more clearly convey emotions than using a general model.
But, arguments can also be made for the contrary:
(1) For example, most robot art systems today are not contingent and creative, but are either controlled by humans or paint independently without humans.
People might prefer a system that is not creative but is just controlled by them due to familiarity with using tools, as familiarity supports usability and technological acceptance, and control supports enjoyment.
Conversely, people might prefer a system which generates nice art regardless of their performance, and can surprise, kindle imagination, stimulate, and inspire.
(2) It is not clear if users know which kinds of visually depicted concepts will best express various emotions to them, due also to bias in their einstellungs, or how to map from general concepts to a concrete visual expression which will evoke a desired emotion.

To provide exploratory insight into these questions, a user study was conducted.
A simplified scenario was adopted in regard to robot strategy, the kind of information used to generate art, the degree of personalization, the kind of art, and target emotions: 
For the robot's art strategy three systems were considered: contingent and creative, only contingent, and only creative.
For the model, self-disclosed information was used.
Personalization was examined in a binary manner (either generic or personalized).
Where possible the generic sketches were used as a basis for personalization since they were artistically attractive and complexity could be a factor for people's preferences; the more aesthetically pleasing and attractive, the stronger participants might feel emotions, as in marketing or advertising.
Likewise, the kind of art was examined in a binary manner, as either abstract or representative.
Since the aim was to get insight into how visual expressions can be personalized and the focus was not on the implementation, paintings were represented using sketches which require less time to produce.
For target emotions, four representative discrete Ekman emotions were selected, one per quadrant in valence-arousal space: happy (high valence, high arousal), relaxed (high valence, low arousal), sad (low valence, low arousal), angry (low valence, high arousal).

Within this context, to gain some exploratory insight into these questions a simplified user study was conducted, using an online survey in two parts: 
The first part acquired data to build a personalized emotional profile, and the second part checked how participants perceived art generated based on their profile using a Wizard of Oz approach.
Thus the first part answers the first question and sets up the answer for the second question, which is provided by the second part of the survey.

\subsubsection{Participants}

Eleven adults from our university (3 female, 8 male; average age: 35.4, SD= 12.2, from 8 countries) participated in the first step of the survey.
The first four participants were then asked to help out in the second part (3 female, 1 male; average age: 30.2, SD=5.0); not all participants were asked to complete the second part due to the time required to generate personalized images.
Participants received no compensation.

\subsubsection{Procedure}

Participants were sent links to a Google Forms survey.
In total the survey took approximately a half-hour to complete, with 20 minutes for the first step and 10 minutes for the second step.

\begin{figure}[h!]
\begin{center}
\includegraphics[width=\linewidth]{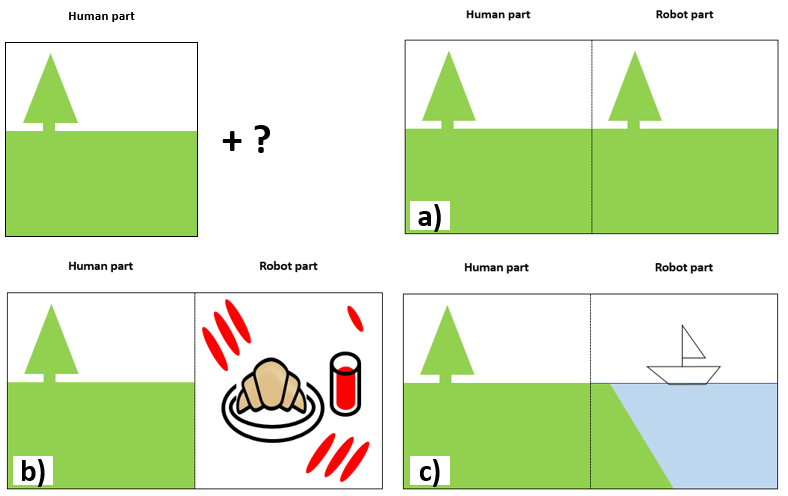}
\end{center}
\caption{Images used to assess how people feel about a robot's art-making strategy: (a) exogenous system (the robot's art is influenced entirely by what the human does), (b) endogenous system (the robot's art is not at all influenced by what the human does), (c) proposed system (the robot's art seeks to express contingency and creativity by maintaining a balance of exogenous and endogenous concerns.}\label{figGenericStrategy}
\end{figure}

In the first step of the survey, participants answered questions about the robot's strategy for making art, then disclosed information about what symbols (representational or abstract) make them feel emotions.
For the robot's strategy the participants were asked to imagine that a human and robot were painting something together, where the human has painted something on the left side, then the robot has painted the part on the right; this had taken place for three different versions of a robot, resulting in three images, as shown in Figure \ref{figGenericStrategy}.
For each image, participants used a 5-point Likert scale to rate three statements:
``The robot's art fits emotionally with the human's art (i.e., the emotions expressed in both seem similar).''
``The robot's art is creative.''
``I would like to do art with this robot.''

In the second half of the first step, participants described which symbols/colors/shapes made them feel happy, relaxed, sad, or angry. The participants were also given a chance to add free comments.

After the first step, the participants' responses were used to generate art, eight personalized images (four abstract and four representational, with one per emotion).
Personalized sketches were generated offline using the Wizard of Oz technique.
The participants' words were input into Google Image Search for images labelled free to reuse. 
To avoid using specific images and communicating unintended signals ``clip art'' was added to the search.
For search queries where no appropriate image was found, synonyms were used.
Images were manually selected to download and assembled into a single composition, then an art program effect was used to make the composition appear like an oil painting.

The eight generic images were created once and used for all participants.
Representational images were generated based on OASIS.
The most extreme four OASIS images were found corresponding to each emotion target.
Abstract images were generated by our artists in line with our previous work based on heuristics including Ståhl's model.
Images were made to appear like oil paintings using an art software.

When the art had been generated, participants conducted the second step of the survey.

This involved assessing 16 images (personalization (yes or no) vs. art type (abstract or representational) vs. four emotions).
First, participants were asked how they felt and if they had knowledge about art.

Then the participants conducted four comparisons, once per emotion, in which they ranked the four images for each emotion (personalized/representational, personalized/abstract, generic/representational, generic/abstract) in order of how much they expressed each emotion (happy, relaxed, sad, angry).
Additionally, as a check to see how the participants perceived individual sketches, the Self-Assessment Manikin (SAM) was used to rate the valence and arousal of each sketch before comparisons.

Image orders were randomized. Each participant received a personalized form for the second part of the survey.
Free answers from the survey were also coded.

\section{Results}

\subsection{Art-making Strategy}

\begin{figure}[h!]
\begin{center}
\includegraphics[width=\linewidth]{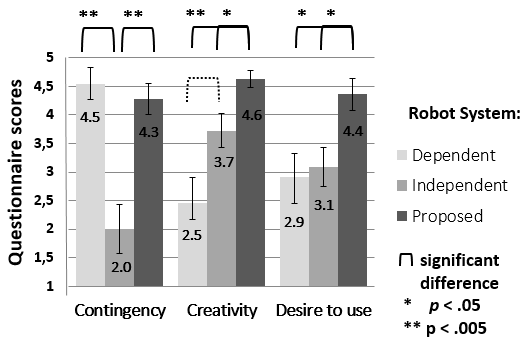}
\end{center}
\caption{Questionnaire results for robot art-making strategy}\label{figQuestionnaire}
\end{figure}

For the first part of the survey, Figure \ref{figQuestionnaire} shows results for how participants perceived the three art-making strategies.

A repeated measures ANOVA indicated that the systems were perceived differently in terms of contingency (F(2, 20) = 26.3, p < .001, $\eta^2$ = .724).
Post-hoc analysis with the Bonferroni adjustment indicated that the independent system was perceived to be less contingent than either of the other systems (p < .001).
No difference was observed between the dependent and proposed systems (p=1,0).

Likewise, the systems were perceived differently in terms of creativity (F(2, 20) = 12,6, p < .001, $\eta^2$ = .558).
Post-hoc analysis with the Bonferroni adjustment indicated that the proposed system was considered more creative than the dependent and independent systems (p= .002 and p<.05), with a trend to significance for the latter two (p= .09). The difference between proposed and independent systems might have been caused by the variation in interpretations of what is creative (e.g., a perception of the proposed system's extension as being more useful) \citep{diedrich2015creative}.

Desire to make art with the robot also differed based on the robot's strategy (F(2, 20) = 7.6, p =.004, $\eta^2$ = .432).
Post-hoc analysis with the Bonferroni adjustment indicated that participants would prefer to make art with the robot which was both contingent and creative, rather than just contingent or just creative (p= .028 and p=.01).
Furthermore, ten out of eleven participants said they would prefer to make art with the proposed system; the one participant who preferred the independent system pointed out that its interpretation was interesting for being largely different from the original.

Thus, the basic premise of this work, that a contingent and creative system might be more desirable than a system which is only contingent or only creative, was supported.

\subsection{Emotional triggers}

Participants' self-disclosures about which symbols elicit emotions are collected in Table \ref{tabSymbols} and Table \ref{tabAbstract}.

\begin{table}
\caption{Typical symbols disclosed by more than one participant as eliciting emotions; typical means described by more than one participant. Numbers beside each coded label indicate the number of mentioning participants. Symbols which elicited more than one kind of emotion are indicated in bold. Comments in parentheses are given for clarification.}
\label{tabSymbols}  
\begin{tabularx}{\linewidth}{ | X | X | X | X | } \hline
Happy & Relaxed & Sad & Angry\\ 
\hline
\\[0.1mm]
\textbf{sports} 6, \textbf{family} 5, \textbf{food and drink} 5, \textbf{nature} 5, traveling 3, \textbf{music} 2, \textbf{work} 2, \textbf{visual leisure activities} 2 & 
\textbf{food and drink} 5, \textbf{visual leisure activities} 5, \textbf{nature} 4, \textbf{sports} (exercise) 3, \textbf{music} 3, rest 2, \textbf{work} (finishing work) 2, washing 2, \textbf{family} 2 & 
\textbf{failure} 6, \textbf{abusiveness} 4, global problems (hunger, poverty, sickness) 4, partings 2, loneliness 2, \textbf{nature} (bad weather) 2, \textbf{injustice} 2, laziness 2 & 
\textbf{abusiveness} 5, \textbf{injustice} 4, stupidity 4, noise/shouting 2, traffic 2, \textbf{failure} 2 \\
\noalign{\smallskip}\hline
\end{tabularx} 
\end{table}

\begin{table}
\caption{Abstract art elements disclosed by participants as eliciting emotions}
\label{tabAbstract}  
\begin{tabularx}{\linewidth}{ | X | X | X | X | X |} \hline
 & Happy & Relaxed & Sad & Angry\\ 
yellow & 		\textbf{6}& 	-& 	1& 	-\\ 
orange & 		3& 	-& 	-& 	1\\ 
pink	& 			3& 	1& 	-& 	1\\ 
purple& 			3& 	1& 	1& 	-\\ 
green&			4&	3&	-&	-\\ 
white&			3&	\textbf{5}&	-&	1\\ 
blue&			3&	\textbf{5}&	1&	-\\ 
black&			1&	-&	\textbf{4}&	3\\ 
red&			1&	-&	2&	\textbf{6}\\ 
brown&			-&	-&	\textbf{4}&	-\\ 
warm colors&	-&	1&	-&	1\\ 
dark colors&		-&	1&	1&	-\\ 
[0.5ex] 
\hdashline
circle \rule{0mm}{5mm}  &			\textbf{6}&	\textbf{6}&	2&	-\\ 
triangle&		2&	1&	3&	\textbf{4}\\ 
square&			2&	-&	3&	1\\ 
horizontal lines&	1&	4&	-&	1\\ 
vertical lines&	1&	3&	3&	-\\ 
diagonal	lines& 	3&	-&	2&	3\\ 
\hline
Total  \rule{0mm}{5mm}&	42&	31&	27&	22\\ 
\noalign{\smallskip}\hline
\end{tabularx} 
\end{table}

In line with previous work, there was a high degree of variation in participants' responses: 243 labels were provided (about half symbols, half abstract), with 49 typical categories mentioned by more than one participant (31 symbols, 18 abstract).

For symbols, 22 out of 31 categories (71\%) were shared between emotions.
Nature was considered to be a cause for either happy, relaxed, or sad emotions.

For colors, out of 12 categories, ten (83\%) were described as eliciting both positive and negative emotions, and eight (67\%) were both calm and aroused
(for valence the only two exceptions were green and brown, which were positive and negative respectively, and orange, pink, brown, and dark colors, where orange and pink were considered aroused and brown and dark colors were considered to be calm). Moreover, considering that red is a warm color, red was described as eliciting every emotion.

For shapes, there was likewise much variation.
Not one typical category was considered by the participants to be either positive or negative; all were labelled as both positive and negative.
As an extreme example, triangles were felt to elicit every emotion.

Along the way it was observed that there were more responses for positive than negative emotions, where some participants mentioned that all shapes seemed positive to them, or gave no responses at all regarding negative emotions.
Also, various references were made to personal information which was not available to the experiments: e.g., ``my programming errors'',  ``my brother'', ``my cat'', ``my child'', ``my mom'', ``my home'', and ``my parents''.

\subsection{Assessment of the sketches}

\begin{figure}[h!]
\begin{center}
\includegraphics[width=\linewidth]{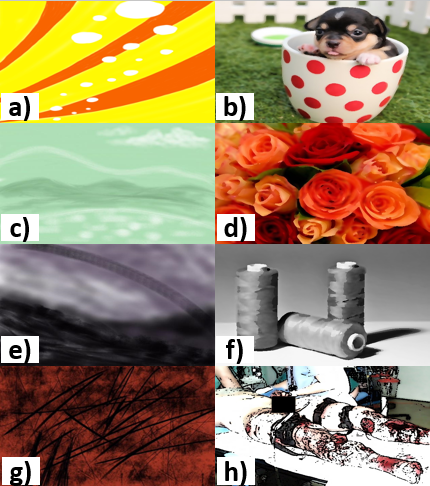}
\end{center}
\caption{Generic sketches: (left) abstract, (right) representational; (a-b) happy, (c-d) relaxed, (e-f) sad, (g-h) angry}\label{figGenericSketches}
\end{figure}

\begin{figure}[h!]
\begin{center}
\includegraphics[width=\linewidth]{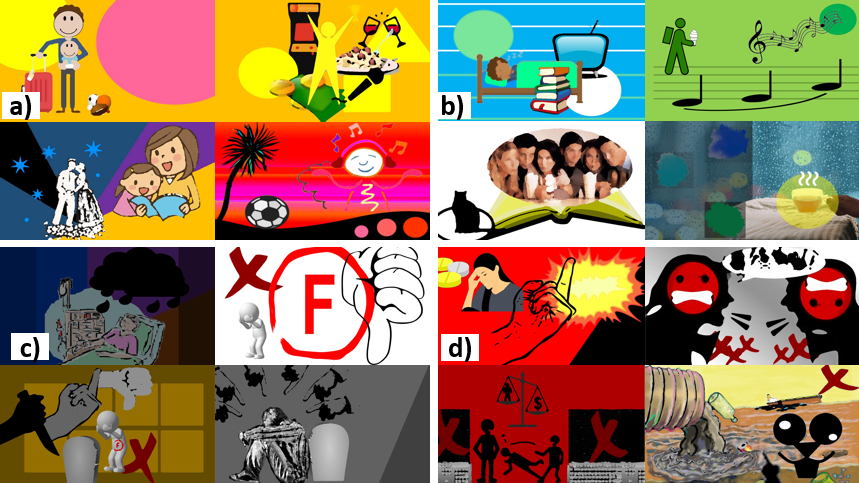}
\end{center}
\caption{Personalized sketches for four participants: (a) happy, (b) relaxed, (c) sad, (d) angry. }\label{figPersonalizedSketches}
\end{figure}

For the second part of the survey, our artists contributed with the generation of sketches.
Figure\ref{figGenericSketches} shows the eight generic sketches used, and Figure\ref{figPersonalizedSketches} shows 32 personalized sketches also generated based on the self-disclosed data from participants.

Data from one participant who described their emotional state at the time of responding as ``depressed'' were not used.
For the SAM assessments, responses were different from those of the other participants: the sketch of a small puppy (selected by the participant as the happiest of the four images intended to express happiness in the comparisons) was rated as highly negative, whereas the generic sketches of an injured bleeding person, as well as the sad and angry generic abstract sketches, were assessed with the most positive score possible.
For the comparisons, a pattern could not be seen in the responses, with a different category considered to best express emotion each time.
Although humans occasionally provide misinformation\citep{clabaugh2017interactive}, the inconsistencies might have been due to feeling depressed, which can involve anhedonia (losing interest, motivation, and pleasure in activities formerly enjoyed) and negative fixation.
This suggests a challenge that a robot might have in painting with someone who is depressed.

The personalized representational art was most frequently described as best conveying the intended emotions (92\%, 11/12, random chance 25\%).
The exception might have arisen from ambiguity; this participant disclosed that warm colors were both happy and angry, and provided abstract concepts like ignorance as examples of stimuli that elicited anger.
The generic representational art was most frequently described as least conveying the intended emotions (44\%, 7/16).
Abstract art, generic and personalized, was in the middle.
The most extreme valence and arousal ratings were also associated with the personalized sketches.
The most positive valence and arousal scores were given for the personalized representational sketches expressing happiness (1.3 and 2.3 on the 9-point scale), lowest valence for the personalized representational sketch expressing sadness (8.7), and lowest arousal for the personalized abstract sketch expressing relaxation (7.5). 
Along the way, the variance in scores for representational and abstract art was also checked. 
For valence, there was more average variance for abstract art than representational art (1.5 vs .85); however, the case was reversed for arousal (1.5 versus 2.2).

It is noted that if the data from the depressed participant are included, the results are similar but noisier, with higher variation and less extreme averages.
The personalized representational art was most frequently described as best conveying the intended emotions (75\%, 12/16); that the participants perceived the systems differently was also confirmed by a Chi-squared test with Yates correction ($\chi^2$(3, N=16) = 17.8, p = .0005).
Also, the most extreme valence and arousal ratings were again associated with the personalized sketches.

\section{Discussion}

In summary, the current article proposed a design for a robot capable of painting with a person, to convey contingency and artistry based on visual metaphors, in a personalized manner.
A simplified user study conducted to gain some extra insight indicated that: 
\begin{itemize}
\item{Participants would prefer to make art with a system which is both contingent and creative.}
\item{Participants' self-disclosed perceptions of art were highly idiosyncratic and varied greatly, in line with observations in previous work.}
\item{Self-disclosure can be used to better communicate emotions to people through representational art.}
\end{itemize}

The results are limited by the exploratory approach in building a simplified prototype and conducting a small user study, but they may provide some opportunity for initial discussion, regarding possible reasons for the observed results, how a richer art-making strategy might be defined, and finally how a more complex user model might look like, as below:

\subsection{Personalized abstract art}

One unexpected result was that personalization seemed to modulate the clarity with which emotions were perceived in representational, but not abstract, sketches.
Three potential causes suggested themselves:
One was that self-disclosed preferences for abstract art often overlapped with general guidelines: e.g., red and black with diagonal lines to represent anger.
By contrast, the space for representational art is much more expansive; for example, no participant mentioned a small puppy in a cup in a grass field (the generic image) as something they would consider to express happiness.
Another potential cause is that the participants might not have known themselves what kind of abstract art would make them feel a certain way.
This was supported by some comments by participants.
One potentially useful approach which could be considered was introduced in a paper by Lee at al., which proposed a reflective approach to personalization intended to empower users by querying to encourage thought about goals before starting an activity \citep{lee2015personalization}; one interesting feature of this work also was that, although robot designers typically try to avoid repetition which could bore humans, interviewed human experts suggested the importance of repeatedly querying users to get at hidden motivations.
Finally, our measurements through a sparse open-ended survey might not have been sufficient to model the participant's preferences.
Aside from merely introducing more questions and considering other factors such as composition, one more complex approach which might yield better results could be to use interactive personalization throughout a more extended period \citep{clabaugh2017interactive}, possibly like the series of questions in an eye exam; similar to the work by Lee et al., despite foreseeing a possibility of survey fatigue, participants in this work reported enjoying being prompted frequently.

\subsection{ Art-making Strategy}

The current work represents a simplified foray into a highly complex area; numerous improvements are likely to be possible by considering this complex phenomenon in more detail and replacing simplified assumptions with richer formulations.

Art is highly complex.
The meaning of art is not always clear to an observer (human or robot) because art is polysemous, overdetermined, and polyvalent.
From the perspective of the artist, the elements of art such as colors are polysemous in the sense that artists will use the same techniques for different reasons: one person might use black because they like the color, another because they want to express a bleak feeling. 
Similarly, art is overdetermined, in the sense that an expression typically arises as from a condensation of multiple causes: a person might use black because they like the color, want to express a bleak feeling, saw a black painting recently, are running out of other colors, and have seen that it is night.
This also relates to emotions; here single emotions were targeted, but emotions in the wild can be much more complex, from the perspective of mixed emotions, referents, timing (dependence on the ``residue'' of recent events), and polysemy.
Moreover, from the perspective of the observer, art is polyvalent; the observer is likely to add their own bias in interpreting, also seeing spurious connections where there are none or nothing where there is a connection (i.e., pareidolia-randomania spectrum).
Additionally, art can contain multiple symbols, possibly nested.

To address such complexities in a human's art, in general, a rich user model (discussed more below), improved modeling of emotions, awareness of context, advanced recognition and inference capabilities, as well as interaction experience, might be helpful for accurately interpreting a human's art. 
Regarding how a robot could seek to deal with multiple symbols, various options exist. If symbols fit well together, like water, palm trees, sand, and a reclining chair, the robot could interpret this as one symbol for a beach. If symbols appear disparate, insight could be taken from an interesting study in Human-Robot Interaction (HRI) that proposed that a robot should perhaps react to both typical patterns in a human's behavior as well as novel behaviors \citep{glas2017personal}. In the current context, this could mean, if a person 
paints a purple background more than once, or paints an uncommon symbol like an aye-aye, that the robot could react to this symbol; additionally, a robot could also focus more on symbols which have an extreme emotional meaning, like skulls, where the cost for ignoring it in a therapy session might be high.

For the robot's art, the robot can also seek to have a model of the distributions of symbols, and how to position symbols on the canvas in relation to the human's art to make intended references clear, given the commingling and contiguity of symbols.
For example, left-facing human faces in paintings are more prevalent (due to the ``left cheek bias'', purportedly arising from better control of the left hemiface by the emotion-dominant right hemisphere), which can be considered in selecting where and what a robot will paint (if the robot paints something on the left, a connection between the person and symbol will likely emerge, affecting the message embedded in the art).
Also, the interaction overall should be considered: art generated through an interaction features a chiasmic relationship; a cycle of entrainment can ensue in which a robot reacts to the human's art, and the human reacts to the robot's art, also over possibly many sessions.
A more complex system could choose different locations to paint, whether the robot's and human's art should fit seamlessly together like a symbolon or if there can be a visible gradient, and whether the robot can modify the human's art, etc.
A robot can also make a judgement of how critical it is to convey meaning clearly. In therapy for a first time user, monosemous symbols could be preferred for clarity, whereas some errors could be allowed in playing with an experienced user.

Another assumption in the current work is that a desirable initial strategy could involve partial matching by a robot that acts as a human's partner in art-making for enjoyment or therapy.
The usefulness of investigating matching, in relation to contingency and empathy, could be perceived in that matching seems to be quite common in human behavior, e.g., in affective mimesis (chameleon effect) and imitation (also prevalent in child-adult interactions).
This does not mean that such a robot would be only capable of supporting venting rather than positive distraction when confronted with negative emotions in therapy. 
Rather, for a first step it was felt that ignoring a person and painting something positive might less challenging of a task than finding an approach to accomplish such partial matching, such that both matching and distraction could be combined, either spatially or temporally.
Moreover, if the eventual goal is to reach the kind of skill that a human therapist might wield, a more complex strategy is required.
For example, in the responsive art paradigm, if a person paints a sad scene of themselves dying, a therapist could also express sadness by painting the person's loved ones grieving.
Or if a person paints themself surrounded by crazy aliens, the therapist could draw them as an alien as well, suggesting that everyone has idiosyncrasies and there might be more in common than is first thought. 
To be able to interact at this level, more advanced robots should be able to perceive not just basic sentiment, but the semantics of messages embedded in art, and be able to come up with meliorative responses.

Regarding the robot's role as a partner, from a means-end theory perspective, users might see a robot as a means to explore, gain self-concept clarity, and achieve; and, typically humanistic therapy also seeks to place users at the center. 
Therefore, a role as an assistant or foil might be more desirable than a fully equal partner.
In other words, a robot could work more on backgrounds and support, to let the human's work shine more.

Regarding the context, future work could also investigate how a robot's behavior might differ if the desired outcome is enjoyment or therapy.
There would appear to be some common area.
For example, humor can be generated by changing one aspect in an otherwise normal scenario in an unexpected way, but this would seem to bear some similarity to our approach for creativity through moving one step away toward the ``adjacent possible'' via a visual metaphor.
A difference might be that for creativity in art, an underlying goal is to point out something relevant which might be overlooked, whereas in humor the goal might be more to stress incongruence and absurdity.
However, the line is not maybe so clear: for example, a very tall hat could appear humorous, but could also be creative if it somehow seeks to bring attention to our assumptions that paintings typically have aspect ratios of 3:4, or that height can express dominance.

Another assumption in the current work is that emotions are a useful property to consider in generating visual metaphors.
Semantic associations, although also possible, could have a higher risk of failing: for example, if a person paints a positive image of a nurse, the robot could paint a negative image of a dying person or a needle; if a person paints red in a sunset, a different meaning would result from painting red (blood) coming from a person.
In short, emotion is used here as a tool to assign a simplified underlying meaning to a visual communication.
However, a more advanced robot that can avoid undesired semantic associations could use such metaphors without considering underlying emotion.

\subsection{User model}

The current article proposed a simplified user model with a small number of open-ended questions, which appears to have been sufficient to allow personalized representational art to convey emotions, but a richer model might perform better.
Also, the proposed approach cannot deal with profile sparseness and the cold-start problem, especially if a user is unwilling or incapable of training the system through self-disclosure.
Below a more complex approach is proposed:

As before, the user model module should accept an emotional state as input and output an image expected to elicit the emotion for a particular person, brute force in obtaining data for every single concept is not possible, and it is desired to identify tricky concepts for which a person's perception differs most from a generic model.

A more sophosticated approach could involve modifying the basic concept of a modular layered architecture to create an adaptive model \citep{hylving2013evolving}.
It could contain both a list of properties that affect emotion perception, as well as two maps between emotions and artistic expression components (one representational, one abstract).
Emotions are typically represented as discrete categories when interacting with people, and as dimensions (e.g., valence and arousal) when working with computations.

The model has three layers or levels: generic, stereotype, and known.
If there is nothing known about a person, the model uses a generic scheme like that used to generate the generic sketches; e.g., like using Ståhl's model for colors.
If there is some basic implicit stereotypical information which can be obtained, like location and gender, an adapted version of the generic model can be used.
Else if rich explicit knowledge is obtained from the user, a caregiver, or other companion such as a robot (e.g., how each color makes them feel), generic or 
implicit values can be replaced and adapted. In the extreme case that there is full explicit knowledge no generic and implicit values are used.
Thus, the model is dynamically updated and can deal with some missing data (implicit or explicit), and explicit knowledge has precedence over implicit and generic 
values; in other words, the model is a hybrid of a stereotype-based user model and a fully personalized adaptive model, depending on the information available.

Regarding the modules of the architecture, these are symbols encoded within the maps.
Various structures could be used  to represent the maps between concepts and emotions and their interrelationships, such as trees or hypergraphs.
For example, a simplified tree structure, following the style of WordNet, could be used, where leaves are actual images associated with a baseline emotion determined empirically by a 
user study (e.g., in valence and arousal), properties such as gender or age, and any known information from the user.
It is not possible to assume a normal distribution, as a category could contain both positive and negative examples, but the average and standard deviation could 
be a useful simplification in selecting a concept to paint.
For example, there could be a few images of golden retrievers, linked to a higher-level category representing hunting dogs, and then above that dogs in 
general. The average emotion for a dog can be calculated as the mean of the leaves below it. Likewise, the expected emotion associated with a new leaf can be 
estimated to be similar to that of the node above it, or to the most similar leaf (e.g., using k Nearest Neighbors). 
Thus, the standard deviation might be low for ``golden retriever'', then high for ``hunting dog'' (which might include images of dead prey and guns), then lower again for 
``dog''; based on this, a robot which wants to minimize the likely difference between the emotions to be conveyed and expressed by the metaphor 
might select a category such as ``golden retriever'' or ``dog'' rather than ``hunting dog''. In other words, the robot can seek to find a good level of concept to 
paint.
A second map could be used to hold data for more abstract concepts, such as colors, and shapes such as lines (horizontal, vertical, diagonal), triangles, and circles.

Thus, the maps are structured as taxonomies/hierarchies. 
The modules of the architecture are symbols encoded in the maps as sub-trees (e.g., ``cat'' and ``dog'', which could possibly be swapped out for one another to generate a visible metaphor).

For the list, variables that could be used for inference include behavior, culture, preferences, age, gender, and personality.
A few examples of how inference could be conducted, for some potential tricky concepts where variance might exist, are given below.
For behavior, attitudes toward animals can vary strongly, such as in people with allergies or traumatic experiences; approaching with a smile or recoiling in fear from an animal could indicate positive or negative emotions. 
An example of how culture can affect emotions perceived in art is how gothic subculture attributes positive characteristics to dark expressions which are 
otherwise often considered to be negative such as black, skulls, or graves.
For preferences, smiling can be seen as positive or negative (e.g., showing empathy or derision); or, one person might like the thrill of action movies, whereas another might detest 
violence.
Regarding age, older people have been reported to prefer colors of shorter rather than longer wavelength (blue, green, violet, vs. red, orange, yellow), although color preferences vary greatly\citep{birren2016color}.
For gender, blue has been associated with masculinity \citep{ellis2001color}, although there has been some opposition to such ``pinkification'', or the indoctrination that children should be forced to use certain colors depending on their gender.
Other properties of interest are also likely to  exist. For example, who is with a person could affect them. A person might not want to admit that they have an emotional reaction to a stimulus in front of a stranger, or find a stimulus more pleasant when seen with a friend.
A person's current or past context might also influence. If a person sees a painting when angry, they might perceive it more negatively.
The same might be the case if a person is cold, wet from rain, tired, or has a cold.

The easiest way to detect tricky concepts could be to directly ask a person, as in the user study, or their caregivers, or acquaintances.
If this is not possible, how could a robot fill in such a model for how people will emotionally react, in some way which doesn't require much effort?
A companion robot which spends much time with a person (e.g., following or walking beside them) could observe reactions to some stimulus, e.g., using a camera to watch facial expressions, gaze tracker, brainwaves, perspiration/galvanic skin response, or open-ended questions in everyday conversation. If a person visits an art museum, the time a person spends in front of a painting could also be used.
Culture and preferences: Preference elicitation can involve checking a person's digital footprint on social media such as Facebook or Instagram; on Facebook 
users can select to react to posts with emoji representing liking, love, mirth, surprise, sadness, or anger (although the thumbs up symbol can be used in 
general, as the referent might not be clear, and there is no guarantee that labels are correct), and the kind of images posted or not posted on Instagram 
could provide insight into a person's social identity, and thereby their ideals, attitudes, and preferences.
Another possibility is for a robot to detect preferences based on what a person chooses to surround themselves with, either at home or in their workplace (and 
what is absent); e.g., object detection could be used by a robot in a person's house or from photos. This can be difficult because, for example, items at an 
office will likely be a mixture of work items, items received from others, and items people like, which might be difficult to tell apart. Another source 
of information could be a person's purchase history. How could knowledge of preference be used to convey other emotions, like anger or sadness? E.g., the 
person's preferred objects could be drawn broken.
A robot could also seek to detect a person's personality by interacting \citep{abe2020estimating}.
If age and gender are not explicitly disclosed to the system, they can be detected with computer vision or sound.
To gain information from a human, a humanoid form might be more effective at obtaining information than an abstract form like a 
phone \citep{jinnai2020mult}.
Of course not all tricky concepts might be easy to detect, like Google's example of a person in a wheelchair chasing a turkey; e.g., if a person hates red jackets 
since their acquaintance was assaulted by someone with a red jacket. 
Thus perfect personalization might not be possible, although this might be acceptable in some contexts.
Such implicit knowledge can be stored in a list.

Thus a system could adapt automatically based on new information. 
For example, a negative reaction to a golden retriever could involve updating the 
emotional ``weights'' attached to nodes for ``golden retriever'', ``hunting dog'', 
``dog'', and ``animal'', via a backpropagation-style update by computing the gradient 
of a loss function with respect to the weights.
Adaptation could also occur on the fly; for example, a robot could start to draw 
a dog, but change to a different animal if a person's emotional reaction is 
negative.

Another natural analogy might be to recommender systems, since the robot 
seeks to find something which would be good to show a human; as such content-
based filtering and collaborative filtering could be applied.
For example, for content-based filtering, if pythons elicit negative emotions in a person, 
then probably vipers will do the same.
For collaborative filtering, if one person thinks lakes and trees are relaxing, 
another person who thinks lakes are relaxing will maybe also find trees to 
be relaxing.  
There are also direct analogies to creativity in desired properties for a 
recommender systems such as ``diversity'' and ``serendipity'' (how surprising 
recommendations are), and ``persistence'' can be related to contingency.

Adaptation could also occur in more complex cases, such as more than one human is 
involved in painting with the robot, to accommodate groups and people from different backgrounds with 
different profiles (child/adult, female/male, etc).
One naive strategy could be to blend weights, e.g., 50-50. However a better 
approach might be to use a symbol expected to minimizes a loss function; for 
example, if a robot is painting for an adult and a child, the robot should likely 
use the more restricted profile of the child to elide showing stimuli which 
could scare or offend.

It could be argued that two maps might not be required.
How people perceive an artistic property like a color can be related to how 
people perceive typical entities with that property. 
For example, on an island with blue prey and red predators, it might be expected that red is seen with less positive valence than blue.
Then, an estimate could be formed by linearly combining emotional values for 
entities with a certain property, possibly with uneven weights.
However, this is just speculation, and it is not clear how to find appropriate weights.

Another concern for a user model can be cross-application interoperability.
Interoperability via a shared/unified model can be difficult in open, dynamic, and exploratory scenarios such as the current one, such that conversion could be 
more practical (e.g., semantic web approaches could be useful).


\subsection{Ethics}

Various potential ethical concerns exist which designers should consider; a few are suggested below.
The robot's choice of what to paint should not be hurtful or embarrassing.
It should be clear that the robot should not be seen as an expert which diagnoses 
a person, but rather as a partner facilitating self-reflection.
It should be the person's decision if and how the robot should be allowed to 
personalize its painting; the process, and which open data are used, should be 
transparent.
Moreover, private data should be securely dealt with to prevent 
facilitating identity theft or discrimination/stigma based on the emotion 
profile.

Furthermore, socially acceptable symbols should be used which are appropriate for 
the context.
For example, affective image datasets can contain images which might be normally 
considered taboo, like sexual images might be associated with high valence, high 
arousal but this would be inappropriate to paint in front of a child; a gaze 
tracker might identify positive emotion based on seeing an attractive individual, 
but this would be embarrassing to paint in front of someone's partner.
Also, negative stereotypes connected to people's insecurities should not be used; 
for example, the robot should not paint a cake in the assumption that an 
overweight person loves food.
A positive capability for transparency would allow the robot to explain why it 
paints something, to ensure the lack of derogatory logic.
Another consideration is the degree to which a robot should always match the 
emotional feeling to a person's and if there is any danger of undue sheltering 
(i.e., related to filter bubbles and echo chambers).

It might feel disagreeable to be tracked and evaluated, like an invasion of personal space.
The robot should also not get in the way, or be perceived as nagging, such that a person might start to ignore it.

By avoiding such ethical concerns, and employing an appropriate art-making strategy and personalization model, this line of research could be useful to achieve robots which can make art together with humans in an enjoyable way; this in turn could potentially facilitate technological acceptance for robots in human spaces, and also eventually provide an opportunity to learn about emotions and creativity, two phenomena which are tightly intertwined in our natures as humans.


\section*{Conflict of Interest Statement}
The author declares that the research was conducted in the absence of any commercial or financial relationships that could be construed as a potential conflict of interest.

\section*{Author Contributions}
Ideation, experimentation, and writing were conducted by the sole author, M. C., who is accountable for the content of the work.

\section*{Funding}
The author received funding from the Swedish Knowledge Foundation (CAISR 2010/0271).

\section*{Acknowledgments}
Thanks to our artists, Dan Koon and Peter Wahlberg, and others who kindly contributed thoughts.

\section*{Data Availability Statement}
The dataset analyzed for this study is not made available, as emotional perceptions could potentially be used to identify individuals. 

\bibliographystyle{frontiersinSCNS_ENG_HUMS} 
\bibliography{personalized_robot_art_arxiv}


\newpage

\end{document}